\documentclass[aps,reprint,pra,superscriptaddress,floats,floatfix,nobibnotes]{revtex4-2}
\usepackage{amsmath}
\usepackage{amsfonts}
\usepackage{amssymb}
\usepackage{graphicx}
\usepackage{color}
\usepackage[colorlinks=true,citecolor=blue,linkcolor=blue,urlcolor=blue]{hyperref}
\usepackage{times}
\usepackage{amstext}
\usepackage{latexsym}
\usepackage{bm}
\usepackage{mathtools}
\usepackage{orcidlink}

\providecommand{\abs}[1]{\lvert #1 \rvert}
\DeclarePairedDelimiter\bra{\langle}{\rvert}
\DeclarePairedDelimiter\ket{\lvert}{\rangle}
\DeclarePairedDelimiterX\braket[2]{\langle}{\rangle}{#1 \delimsize\vert #2}

\begin{document}

\title{Beam splitter for dark and bright states of light}

\author{Luiz O. R. Solak\,\orcidlink{0000-0002-4760-3357}}
\affiliation{Departamento de Física, Universidade Federal de São Carlos, 13565-905 São Carlos, São Paulo, Brazil}
\affiliation{CESQ/ISIS (UMR 7006), Université de Strasbourg and CNRS, 67000 Strasbourg, France}
\author{Celso J. Villas-Boas\,\orcidlink{0000-0001-5622-786X}}
\affiliation{Departamento de Física, Universidade Federal de São Carlos, 13565-905 São Carlos, São Paulo, Brazil}
\author{Daniel Z. Rossatto\,\orcidlink{0000-0001-9432-1603}}
\email{dz.rossatto@unesp.br}
\affiliation{Universidade Estadual Paulista (UNESP), Instituto de Ciências e Engenharia, 18409-010 Itapeva, São Paulo, Brazil}

\begin{abstract}
Beam splitters are key elements in optical and photonic systems, and are therefore employed in both classical and quantum technologies. Depending on the intended application, these devices can split incident light according to its power, polarization state, or wavelength. In this work, we theoretically present a novel type of beam splitter capable of separating a light beam into its two-mode bright and dark components. We propose a prototype based on an optical cross-cavity system resonantly coupled to a $\Lambda$-type three-level atom. The dark component of the incoming light is transmitted because the antisymmetric collective mode of the cavity setup is decoupled from the atom. Meanwhile, the bright component is reflected due to a quantum interference, which arises from the high cooperativity between the atom and the symmetric collective mode of the cavity setup. Although the device requires only a two-level atom to operate, using a three-level atom allows the device to be turned on or off by controlling the atomic ground states. Our results pave the way for new applications of beam splitters that leverage the collective properties of light. Manipulating and exploiting this additional degree of freedom can advance the field of quantum optics and contribute to the development of quantum technologies.
\end{abstract}

\maketitle

\section{Introduction} \label{sec:1}
In quantum optics, the bright and dark states of matter, more commonly referred to as superradiant and subradiant states~\cite{PhysRev.93.99}, have been extensively investigated over the past decades. Matter in a bright state interact with radiation, absorbing and emitting photons, which leads to typical light-matter interactions and their applications. In contrast, matter in a dark state is invisible to light due to a quantum destructive interference (coherent population trapping) \cite{Arimondo1996}. This state has been explored in atomic \cite{RevModPhys.75.457}, trapped-ion \cite{PhysRevLett.119.220501}, solid-state \cite{PhysRevLett.97.247401,Xu2008,Togan2011,Kumar2016} and hybrid systems \cite{Zhu2014,Zhang2015}, constituting the fundamental basis for a wide range of applications \cite{fleisch,Mucke_2010,RevModPhys.70.707,RevModPhys.87.637,RevModPhys.89.015006,PhysRevA.65.022314,Bajcsy2003,PhysRevLett.84.5094,Ronagel2016,PhysRevA.91.042116}.

More recently, the concept of bright and dark states has been extended to the context of bosonic modes \cite{Delanty2011,Dong2012,PhysRevLett.122.253603,parke2024phononic,PhysRevLett.108.153603,PhysRevLett.110.233602,Gentry2014,PhysRevA.102.023707,PhysRevA.106.013526}, which can also be in states that are coupled or decoupled from matter. It is important to emphasize that the single-mode case exhibits a unique dark state (the vacuum state), whereas the multimode case features an infinite family of dark states with arbitrary numbers of photons \cite{Delanty2011,Mximo2021}. These collective properties of light have been shown to play a crucial role in understanding fundamental principles of physics \cite{Mximo2021}, improving the implementation of quantum tasks \cite{Solak2024}, and paving the way for new quantum devices, such as the one introduced here.

\begin{figure*}[t]
\includegraphics[trim = 0mm 2mm 0mm 0mm,clip,width =\linewidth]{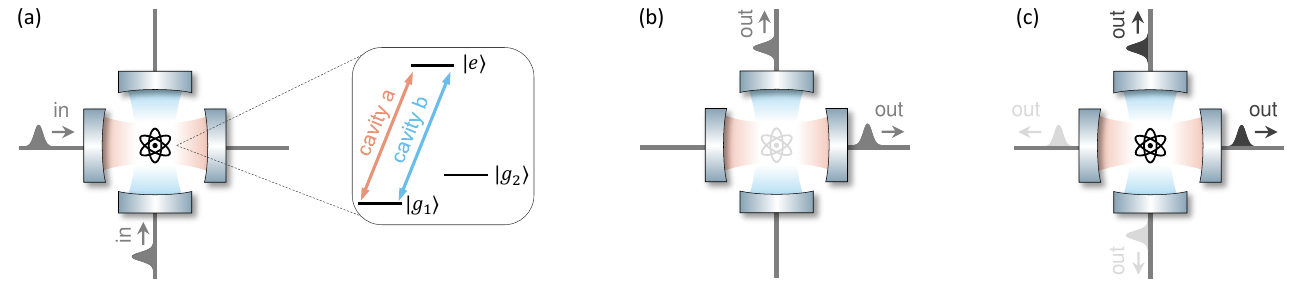}
\caption{\label{setup}(a) A $\Lambda$-type three-level atom resonantly coupled to a system of crossed cavities, comprised by a pair of symmetrical two-sided cavities, upon which a light pulse impinges.
(b) When the atom is in $\ket{g_2}$, it is decoupled from the cavities (empty-cavity scenario), thus the system works as a passthrough device since the incoming field is totally transmitted. (c) In contrast, when the atom is in the state $\ket{g_1}$, a quantum interference due to a high cooperativity causes the system to function as a device that separates a two-mode light into its bright (reflected) and dark (transmitted) components.}
\end{figure*}

In this work, we present a novel kind of beam splitter that separates an incident light beam into its collective two-mode bright (reflected) and dark (transmitted) components. We consider a cross-cavity setup \cite{brekenfeld}, composed of a pair of symmetrical two-sided cavities, both resonantly coupled to the transition $\ket{g_1} \leftrightarrow \ket{e}$ of a $\Lambda$-type three-level atom (two metastable ground states, $\ket{g_1}$ and $\ket{g_2}$, and an excited one $\ket{e}$), as illustrated in Fig.~\ref{setup}(a). Since the atom remains uncoupled from the cavities when it is in $\ket{g_2}$ [Fig.~\ref{setup}(b)], the system works as a passthrough device with the incoming field completely transmitted (empty-cavity scenario). Conversely, when the atom populates $\ket{g_1}$ [Fig.~\ref{setup}(c)], a quantum interference due to a high cooperativity between the atom and the cavity system causes the bright component of an incident light to be reflected, while the dark component continues to be transmitted because it does not interact with the atom. It is worth to stress that using a two-level atom is sufficient to separate the light into its bright and dark components. This implies that these results are not limited to the optical domain and atomic systems~\cite{brekenfeld}, but can also be adapted and extended to solid-state-based systems, such as circuit~\cite{RevModPhys.93.025005} and waveguide quantum electrodynamics~\cite{RevModPhys.95.015002}. However, employing a three-level atom allows the device to be turned on or off by manipulating the atomic state.

Beam splitters are essential devices in both classical and quantum optics, widely used in various optical and photonic systems, as well as in numerous applications in quantum technologies \cite{Makarov2022, Shen2022}. Different variants of beam splitters have been engineered; for example, these devices can split incident light according to its power, polarization state, or wavelength \cite{Shen2022}. Our findings open a new avenue for the application of beam splitters by leveraging the collective properties of light, which provide an additional degree of freedom to be manipulated and exploited.

This paper is organized as follows. Section~\ref{sec:2} provides the model for the open quantum system that we considered. Our main results are discussed in Sec.~\ref{sec:3}, while Sec.~\ref{sec:4} presents our conclusions.

\section{Model} \label{sec:2}
Working in an interaction picture rotating at the resonant frequency of the cavities ($\omega_c$) and under the assumption of the \textit{white-noise limit} \cite{qnoise}, the dynamics of the setup depicted in Fig.~\ref{setup}(a) is governed by ($\hbar = 1$)
\begin{widetext}
    \begin{align}       \label{Hamiltonian}
    H &= \sum_{\ell = r, t} \left\{\int^{\infty}_{-\infty}d\omega \omega A^{\dagger}_{\ell}(\omega) A_{\ell}(\omega) + \int^{\infty}_{-\infty}d\omega \omega B^{\dagger}_{\ell}(\omega) B_{\ell}(\omega) \right\}  + \int^{\infty}_{-\infty}d\omega \omega C^{\dagger}(\omega) C(\omega) + \int^{\infty}_{-\infty}d\omega \omega D^{\dagger}(\omega) D(\omega) \nonumber\\
    &+ \sum_{\ell = r, t} \left\{{\frac{i}{\sqrt{2\pi}}}\int^{\infty}_{-\infty}d\omega\sqrt{2\kappa_{a}^{\ell}}\left[a^{\dagger}A_{\ell}(\omega)-a A^{\dagger}_{\ell}(\omega)\right] + {\frac{i}{\sqrt{2\pi}}}\int^{\infty}_{-\infty}d\omega\sqrt{2\kappa_{b}^{\ell}}\left[b^{\dagger}B_{\ell}(\omega)-b B^{\dagger}_{\ell}(\omega)\right] \right\}\nonumber\\
    &+ {\frac{i}{\sqrt{2\pi}}}\int^{\infty}_{-\infty}d\omega\sqrt{2\Gamma_{1}}\left[C(\omega)\sigma^{1}_{+} - C^{\dagger}(\omega)\sigma^{1}_{-}\right] + {\frac{i}{\sqrt{2\pi}}}\int^{\infty}_{-\infty}d\omega\sqrt{2\Gamma_{2}}\left[D(\omega)\sigma^{2}_{+} - D^{\dagger}(\omega)\sigma^{2}_{-}\right] \nonumber\\
    &+ \underbrace{(g_{a}a+g_{b}b)\sigma_{+}^{1} + (g^{*}_{a}a^{\dagger}+g_{b}^{*}b^{\dagger})\sigma_{-}^{1}}_{H_\text{sys}},
    \end{align}
\end{widetext}
in which $\sigma_{+}^{j}=\ket{e}\bra{g_{j}}$ and $\sigma_{-}^{j}=\ket{g_{j}}\bra{e}$ are atomic ladder operators ($j=1, 2$), $a$ and $b$ ($a^{\dagger}$ and $b^{\dagger}$) are the annihilation (creation) operators of the intracavity modes, while $A_{\ell}(\omega)$ and $B_{\ell}(\omega)$ [$A^{\dagger}_{\ell}(\omega)$ and $B^{\dagger}_{\ell}(\omega)$] are the frequency-dependent annihilation (creation) operators~\cite{Loudon2000} of the bosonic reservoirs of cavities $a$ and $b$, respectively, for $\ell = \{r, t\}$. The index $r$ is associated with the bosonic reservoirs through which the input pulse approaches the cavities and may be reflected subsequently, while the index $t$ is related to the bosonic reservoirs through which the pulse can be transmitted. The decay rates of the field amplitudes of cavities $a$ and $b$ due to the reservoir $\ell$ are represented by $\kappa_{a}^{\ell}$ and $\kappa_{b}^{\ell}$, respectively. Similarly, atomic reservoirs are described by frequency-dependent bosonic operators $C(\omega)$ and $D(\omega)$, which are responsible for atomic decay from $\ket{e}$ to $\ket{g_{1}}$ and to $\ket{g_{2}}$, with rates $\Gamma_{1}$ and $\Gamma_{2}$, respectively. Finally, the Jaynes-Cummings-type coupling strengths between the atomic transition $\ket{g_1}\leftrightarrow\ket{e}$ and the intracavity modes are given by $g_a$ and $g_b$.

Setting 
$g_a=g_b \equiv g \in \mathbb{R}$ and $\kappa_a^{\ell} = \kappa_b^{\ell} \equiv \kappa_{\ell}$ (identical cavities), the system response can be examined through the Heisenberg-Langevin equations (Appendix \ref{app:hl}) \cite{qnoise},
\begin{align}
           &\dot{a}(t) = -ig\sigma_{-}^{1}(t) - (\kappa_r + \kappa_t) a(t) + \sqrt{2\kappa_r} a^{r}_\text{in}(t) \label{adot2}, \\
        &\dot{b}(t) = -ig\sigma_{-}^{1}(t) - (\kappa_r + \kappa_t)b(t) + \sqrt{2\kappa_r} b^r_\text{in}(t) \label{bdot2}, \\
        &\dot{\sigma}_{-}^{1}(t) = ig[ a(t) +  b(t)]\sigma_{z}^{1}(t) - (\Gamma_1 + \Gamma_2)\sigma_{-}^{1}(t), \label{sdot2}
\end{align}
combined with the input-output relation~\cite{walls2008}
\begin{equation}
    z^{\ell}_\text{out}(t) = \sqrt{2\kappa_{\ell}} \, z(t) - z^{\ell}_\text{in}(t), \label{inoutrel}
\end{equation}
in which~\cite{sign}
\begin{align}
    z^{\ell}_\text{out}(t) &=  \dfrac{(-1)^{\delta_{t,\ell}}}{\sqrt{2\pi}} \int^{\infty}_{-\infty}d\omega e^{-i \omega(t-t_\text{out})} \underbrace{Z_{\ell}(\omega, t_\text{out})}_{z^{\ell}_\text{out}(\omega)}, \label{outop} \\
    z^{\ell}_\text{in}(t) &=   \dfrac{(-1)^{\delta_{r,\ell}}}{\sqrt{2\pi}} \int^{\infty}_{-\infty}d\omega e^{-i \omega(t-t_\text{in})} \underbrace{Z_{\ell}(\omega, t_\text{in})}_{z^{\ell}_\text{in}(\omega)}, \label{inop}
\end{align}
for $\{z=a, Z=A\}$ and $\{z=b, Z=B\}$, with $\ell = \{r, t\}$. 
Without loss of generality, the input operators for the atomic reservoirs were suppressed because they are assumed to be initially in the vacuum state. For the same reason, $a^{t}_\text{in} = 0$ and $b^{t}_\text{in} = 0$.

\section{Results} \label{sec:3}
An analytical input-output relation for field amplitudes in the frequency domain can be derived by using the Fourier transform when assuming the Holstein-Primakoff approximation~\cite{holsteinprimakoff} ($\sigma_z = \sigma_{+}^{1}\sigma_{-}^{1} - \sigma_{-}^{1}\sigma_{+}^{1} \approx -\mathbf{1}$, which holds as long as the atom is rarely excited or if the atom is in $\ket{g_2} \leftrightarrow g=0$). At this point, it is convenient to define the collective-mode operators for the incoming field, $X_\text{in}^{\pm}(\omega) = [a^r_\text{in}(\omega) \pm b^r_\text{in}(\omega)]/\sqrt{2}$, for the reflected field, $X_\text{out}^{\pm}(\omega) = [a^r_\text{out}(\omega) \pm b^r_\text{out}(\omega)]/\sqrt{2}$, and for the transmitted one, $Y_\text{out}^{\pm}(\omega) = [a^t_\text{out}(\omega) \pm b^t_\text{out}(\omega)]/\sqrt{2}$.
Considering symmetrical cavities ($\kappa_r = \kappa_t = \kappa/2$), the system response for incoming resonant fields ($\omega = \omega_c$) is given by (Appendix \ref{app:hl})
\begin{align}\label{eqax}
        &X_\text{out}^{\pm}(\omega_c) = x_{\pm}(\omega_c) X_\text{in}^{\pm}(\omega_c),  \\ \label{eqax2}
        &Y_\text{out}^{\pm}(\omega_c) = y_{\pm}(\omega_c) X_\text{in}^{\pm}(\omega_c),   
    \end{align}
with
    \begin{align}
        &x_{+}(\omega_c) = -\frac{C}{1+C} \ket{g_1}\bra{g_1},\\
        &x_{-}(\omega_c) = 0,\\
        &y_{+}(\omega_c) = \frac{1}{1+C}\ket{g_1}\bra{g_1} + \ket{g_2}\bra{g_2},\\
        &y_{-}(\omega_c) = 1.
    \end{align}
in which $C = (g\sqrt{2})^2/\kappa\Gamma$ (cooperativity) and $\Gamma = \Gamma_1 + \Gamma_2$.

Antisymmetric collective mode operators are always decoupled from the atom, as their relations do not depend on $g$. Because of that, in analogy to Dicke states~\cite{PhysRev.93.99} regarding subradiance in multiatom systems, the vacuum state and the states of the type $\ket{\Psi_\mathcal{D}^N}_r^\text{in} = \tfrac{(X_\text{in}^{-})^{\dagger N}}{\sqrt{N !}}\ket{0}\ket{0}$ can be called the \textit{dark states} of the incoming light with $N$ photons~\cite{Delanty2011,Mximo2021}, since the atom (matter) cannot ``see'' the light in these states; the system dynamics in this case is equivalent to the empty-cavity scenario. Similarly, the states $\ket{\Psi_\mathcal{B}^N}_r^\text{in} = \tfrac{(X_\text{in}^{+})^{\dagger N}}{\sqrt{N !}}\ket{0}\ket{0}$ can be called the \textit{bright states} of the incoming light, since the atom can ``see'' the light. Hence, $x_\pm$ and $y_\pm$ mean the complex reflection and transmission coefficients (dependent on the atomic state) for each excitation of the dark ($-$) and bright ($+$) components of the incoming field. 

Note that the incoming field is completely transmitted when the atom is in $\ket{g_2}$, since in this case $x_+ = x_- = 0$ and $y_+ = y_- = 1$ [Fig.~\ref{setup}(b)]. In other words, the system works out as a passthrough device ($\ket{g_2}\ket{\psi}_r^\text{in} \to \ket{g_2}\ket{\psi}_t^\text{out}$). On the other hand, when the atom is in $\ket{g_1}$, the projection of the incoming field into the dark states continues to be transmitted ($x_{-} = 0$ and $y_{-} = 1$), but the projection into the bright states is both partially reflected and transmitted ($x_{+}, y_{+} \ne 0$). However, in a high-cooperativity regime ($C\gg 1$) with respect to the atom and the intracavity symmetric collective mode, the projection of the incoming field into the bright states undergoes a total reflection ($x_{+}\to 1$ and $y_{+} \to 0$). Therefore, in this regime, the system works as a device that separates a two-mode light into its bright (reflected) and dark (transmitted) components [Fig.~\ref{setup}(c)].

For example, let us consider as the input field a single-photon pulse resonantly impinging upon the cavity $a$ in the high-cooperativity regime, such that the initial state of the reservoirs reads 
\begin{align}
    \ket{\psi}_r^\text{in} &= \ket{1_\text{in}}_{\alpha}^{r} \ket{0}_{\beta}^{r}  \ket{0}_{\alpha}^{t}  \ket{0}_{\beta}^{t} \nonumber \\
    &=  [a_\text{in}^r (\omega_c)]^\dagger \ket{0}_{\alpha}^{r} \ket{0}_{\beta}^{r}  \ket{0}_{\alpha}^{t}  \ket{0}_{\beta}^{t} \nonumber\\
    &= \tfrac{1}{\sqrt{2}}\{[X_\text{in}^{-}(\omega_c)]^\dagger + [X_\text{in}^{+}(\omega_c)]^\dagger\}\ket{0}_{\alpha}^{r} \ket{0}_{\beta}^{r}  \ket{0}_{\alpha}^{t}  \ket{0}_{\beta}^{t}\nonumber\\
    &= \tfrac{1}{\sqrt{2}} ( \ket{\Psi_\mathcal{B}^1}_r^\text{in} + \ket{\Psi_\mathcal{D}^1}_r^\text{in})  \ket{0}_{\alpha}^{t}  \ket{0}_{\beta}^{t}, \label{psi_in_r}
\end{align}
with $\ket{0}_{\alpha}^{r}$ and $\ket{0}_{\alpha}^{t}$ representing the reservoirs of cavity $a$ in the vacuum state, in which the input field can be reflected and transmitted, respectively. Furthermore, $\ket{0}_{\beta}^{r}$ and $\ket{0}_{\beta}^{t}$ are defined similarly for the reservoirs of cavity $b$, while $\ket{1_\text{(in)out}}_{\alpha}^{\ell} \equiv [a_\text{(in)out}^{\ell}(\omega_c)]^\dagger\ket{0}_{\alpha}^{\ell}$ and $\ket{1_\text{(in)out}}_{\beta}^{\ell} \equiv [b_\text{(in)out}^{\ell}(\omega_c)]^\dagger\ket{0}_{\beta}^{\ell}$, with $\ell = \{r, t\}$.
In this scenario, our system yields $\ket{g_1}\ket{\psi}_r^\text{in} \to\ket{g_1}\ket{\phi}_{rt}^\text{out}$, with
\begin{align}
    \ket{\phi}_{rt}^\text{out} &= \tfrac{1}{\sqrt{2}}\{[Y_\text{out}^{-}(\omega_c)]^\dagger-[X_\text{out}^{+}(\omega_c)]^\dagger\}\ket{0}_{\alpha}^{r} \ket{0}_{\beta}^{r}  \ket{0}_{\alpha}^{t}  \ket{0}_{\beta}^{t}\nonumber \\ 
    &= \tfrac{1}{\sqrt{2}} (\ket{0}_{\alpha}^{r}  \ket{0}_{\beta}^{r}  \ket{\Psi_\mathcal{D}^1}_t^\text{out}- \ket{\Psi_\mathcal{B}^1}_r^\text{out}  \ket{0}_{\alpha}^{t}  \ket{0}_{\beta}^{t}), \label{psi_out}
\end{align}
in which 
\begin{align}
    \ket{\Psi_\mathcal{B}^1}_r^\text{out} &= [X_\text{out}^{+}(\omega_c)]^\dagger \ket{0}_{\alpha}^{r} \ket{0}_{\beta}^{r} \nonumber \\ &= \tfrac{1}{\sqrt{2}}(\ket{1_\text{out}}_{\alpha}^{r} \ket{0}_{\beta}^{r} + \ket{0}_{\alpha}^{r} \ket{1_\text{out}}_{\beta}^{r}), \\
    \ket{\Psi_\mathcal{D}^1}_t^\text{out} &= [Y_\text{out}^{-}(\omega_c)]^\dagger \ket{0}_{\alpha}^{t} \ket{0}_{\beta}^{t} \nonumber \\
    &= \tfrac{1}{\sqrt{2}}(\ket{1_\text{out}}_{\alpha}^{t} \ket{0}_{\beta}^{t} - \ket{0}_{\alpha}^{t} \ket{1_\text{out}}_{\beta}^{t})
\end{align}
are the reflected and transmitted fields in a single-excitation bright and dark states, respectively.
%
%
%
From Eqs.~\eqref{psi_in_r} and \eqref{psi_out} it is straightforward to notice the splitting of the input field into its bright (reflected) and dark (transmitted) parts, which in this case represents a single-step generation of a four-qubit $W$ state \cite{PhysRevA.62.062314} among the ports of the cavity system, a key ingredient for quantum communication and computing protocols.

With this beam splitter for dark and bright states of light, one can play with different types of input state, also manipulating and measuring the atomic state, to generate diverse interesting output light states. The incident light may also be classical, e.g., two coherent fields with opposite phases decompose in terms of dark states only $\ket{\alpha,-\alpha} = e^{-|\alpha|^2}\sum_{N=0}^{\infty} \tfrac{\alpha^N}{\sqrt{N!}}\ket{\Psi_\mathcal{D}^N}$, while in-phase coherent states decompose exclusively in the bright-state subspace $\ket{\alpha,\alpha} = e^{-|\alpha|^2}\sum_{N=0}^{\infty} \tfrac{\alpha^N}{\sqrt{N!}}\ket{\Psi_\mathcal{B}^N}$ \cite{Mximo2021,parke2024phononic}.

Although Eqs.~\eqref{eqax} and \eqref{eqax2} were obtained imposing the Holstein-Primakoff approximation, their validity can be exactly checked in the Schrödinger picture when the initial state contains only a single excitation in the input field
\begin{align}
    \label{ISa}        \vert \Psi_{t_0}(\omega) \rangle&=\underbrace{\left(\lambda_{1}\ket{g_1}+\lambda_{2}\ket{g_2}\right)}_{\text{Atom}}   \underbrace{(\ket{0}_{a}   \ket{0}_{b})}_{\text{Cavities}} \nonumber\\
    & \times \underbrace{\left(\mu_{a}\ket{1}_{\alpha}^{r} \ket{0}_{\beta}^{r}  + \mu_{b}\ket{0}_{\alpha}^{r} \ket{1}_{\beta}^{r}   \right)\ket{0}_{\alpha}^{t}  \ket{0}_{\beta}^{t}}_{\text{Reservoirs}},
\end{align}
in which $\ket{1}_{\alpha}^{
r}=\int^{\infty}_{-\infty}d\omega \xi_\text{in}(\omega)A^{\dagger}_{r}(\omega)\ket{0}_{\alpha}$ and $\ket{1}_{\beta}^{
r}=\int^{\infty}_{-\infty}d\omega \zeta_\text{in}(\omega)B^{\dagger}_{r}(\omega)\ket{0}_{\beta}$~\cite{Loudon2000}. The square-normalized temporal shapes of the incoming pulses, $\alpha_\text{in}(t)$ and $\beta_\text{in}(t)$, are the Fourier transform of the spectral density functions, $\xi_\text{in}(\omega)$ and $\zeta_\text{in}(\omega)$.

The dynamics can be described by the non-Hermitian Schrödinger equation $i\partial_{t}\ket{\Psi_t}=(H - i\Gamma \sigma_{ee})\ket{\Psi_t}$, where the operators $C(\omega)$ and $D(\omega)$ must be removed from $H$ since the occurrences of photon loss due to spontaneous atomic emission are taken into account through the damping term $i\Gamma \sigma_{ee}$, with $\sigma_{ee} = \ket{e}\bra{e}$. The output pulses that emerge from the cavity $a$ and $b$, towards the reservoir $\ell = \{r,t\}$ and conditioned to the atomic state $\ket{g_k}$ ($k = 1,2$), are described by their temporal shapes $\alpha^{k\ell}_\text{out}(t)$ and $\beta^{k\ell}_\text{out}(t)$, respectively. They are determined by the input-output relations (for identical symmetrical cavities) $\alpha^{k \ell}_\text{out}(t)=\sqrt{\kappa}c_{a}^{k}(t) - \delta_{r,\ell}\alpha^{k \ell}_\text{in}(t)$ and $\beta^{k \ell}_\text{out}(t)=\sqrt{\kappa}c_{b}^{k}(t) - \delta_{r,\ell}\beta^{k \ell}_\text{in}(t)$, with $c_{a}^{k}$ and $c_{b}^{k}$ representing the probability amplitudes of finding a single excitation inside the cavity $a$ and $b$, respectively, when the atom is in $\ket{g_k}$ and the other modes are in the vacuum state (Appendix \ref{app:sch}). For the initial state given in Eq.~\eqref{ISa}, $\alpha^{kr}_\text{in}(t) = \lambda_k \mu_a \alpha_\text{in}(t)$ and $\beta^{kr}_\text{in}(t) = \lambda_k \mu_b \beta_\text{in}(t)$.
Here, the incoming pulse will be described by a single-photon Gaussian wave packet, $\alpha_\text{in}(t) \text{ or } \beta_\text{in}(t)= (\eta\sqrt{\pi})^{-\frac{1}{2}}\exp^{-\frac{1}{2}\frac{(t-t_{0})^2}{\eta^2}}$, in which $t_0$ is the time when its maximum reaches the respective cavity and $\tau_{p}=2\eta\sqrt{2\ln(2)}$ indicates the pulse duration.

\begin{figure}[tb]
\includegraphics[trim = 6.5mm 6.5mm 11.7mm 12mm, width =0.95\linewidth, clip]{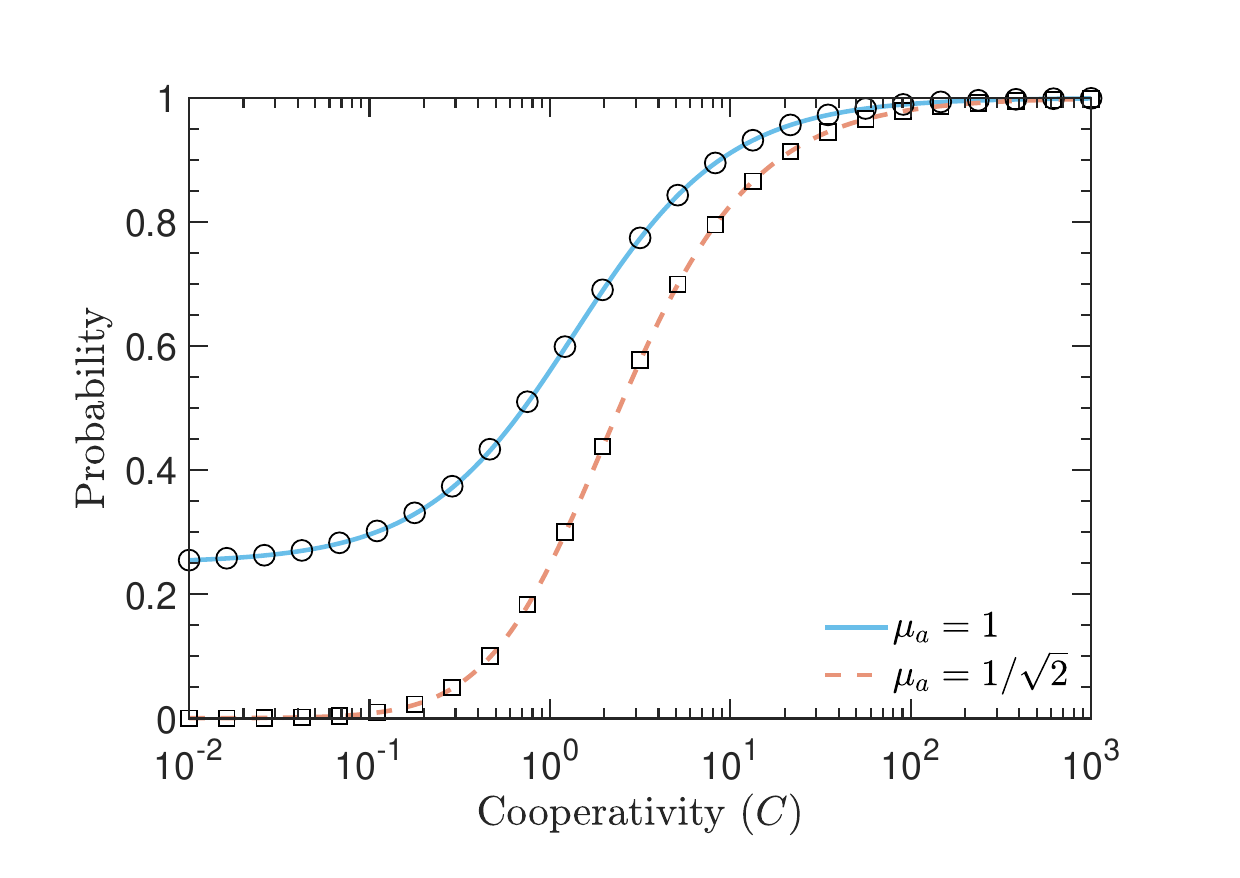}
\caption{\label{comp} Probability of finding the output pulse in $\ket{\psi_\text{tgt}(\omega)}$ as a function of the cooperativity, using  $\lambda_1 =1$, $\Gamma = 0.1\kappa$ and $\kappa \tau_p = 100$. The lines are obtained from the exact Schrödinger equation, while the symbols from the Holstein-Primakoff approximation.}
\end{figure}

As long as $\kappa \tau_p \gg 1$, the Schrödinger approach provides the same exact results given by Eqs.~\eqref{eqax} and \eqref{eqax2} when we have $\lambda_1 = 0$ or $\mu_a = - \mu_b = 1/\sqrt{2}$ in Eq.~\eqref{ISa}. Then, it suffices to analyze the cases with $\lambda_1 = 1$ and $\mu_a \ne - \mu_b$ to check the validity of the Holstein-Primakoff approximation. For this analysis, we consider the atom in $\ket{g_1}$ ($\lambda_1 = 1$), $\xi_\text{in}(\omega) = \zeta_\text{in}(\omega)$ [$\alpha_\text{in}(t) = \beta_\text{in}(t)$], $g_a=g_b \equiv g$ and $\kappa_a^{\ell} = \kappa_b^{\ell} \equiv \kappa/2$. In this case, in a high-cooperativity regime, the reservoirs are expected to be found in the following final state
\begin{align} \label{finalstate}
   \ket{\psi_\text{tgt}(\omega)} &=\tfrac{\mu_{-}}{\sqrt{2}}[A^{\dagger}_{t}(\omega) - B^{\dagger}_{t}(\omega)]\ket{0}_{\alpha}^{r} \ket{0}_{\beta}^{r}  \ket{0}_{\alpha}^{t}\ket{0}_{\beta}^{t}\nonumber \\ &-\tfrac{\mu_{+}}{\sqrt{2}}[A^{\dagger}_{r}(\omega) + B^{\dagger}_{r}(\omega)] \ket{0}_{\alpha}^{r}\ket{0}_{\beta}^{r}\ket{0}_{\alpha}^{t} \ket{0}_{\beta}^{t} ,
\end{align}
with $\mu_{\pm} = (\mu_a \pm \mu_b)/\sqrt{2}$. The probability of the reservoirs reaching $\ket{\psi_\text{tgt}(\omega)}$ for any cooperativity regime is given by
\begin{equation}
    P_\text{tgt}^\text{HP} = \left(\vert\mu_{-}\vert^{2} + \vert\mu_{+}\vert^{2}\frac{C}{1+C} \right)^2,
\end{equation}
according to the Holstein-Primakoff approximation, while
\begin{equation}
    P_\text{tgt} = \int^{+\infty}_{-\infty} \!\!\!\!\!\!\! dt \, \lvert \mu_{-} [\alpha^{1 t}_\text{out}(t) - \beta^{1 t}_\text{out}(t)] - \mu_{+}[\alpha^{1 r}_\text{out}(t) + \beta^{1 r}_\text{out}(t)] \rvert^2,
\end{equation}
according to the exact Schrödinger equation (Appendix \ref{app:prob}).
In Fig.~\ref{comp}, considering $\Gamma = 0.1\kappa$ and $\kappa \tau_p = 100$, we show $P_\text{tgt}^\text{HP}$ (symbols) and $P_\text{tgt}$ (lines) as functions of the cooperativity for two cases, when $\mu_a = 1$ [incoming field in a superposition of bright and dark single-photon states, as in Eq.~\eqref{psi_in_r}] and when $\mu_a = \mu_b = 1/\sqrt{2}$ (incoming field in a single-photon bright state). We observe that both treatments yield the same results for the parameters under consideration, and the probability of finding the output pulse in $\ket{\psi_\text{tgt}(\omega)}$ approaches unity as $C$ increases. For this case, we obtain a probability greater than 90\% for $C=20$ ($g = \kappa$), which is achievable with current technology~\cite{brekenfeld}. Although we use $\Gamma = 0.1\kappa$ in Fig.~\ref{comp}, this result is valid for any value of $\Gamma$, provided that $\tau_p$ is sufficiently long, indicating that the beam splitter does not require strong coupling. Finally, efficiency can be improved by using an ensemble of $N$ identical atoms rather than just a single atom coupled to the cross-cavity setup, because in this case $g \to g_\text{eff} = g\sqrt{N}$ due to the atomic collective effect.

\section{Conclusions} \label{sec:4}
We have theoretically introduced a beam splitter for dark and bright states of light. For its implementation, we consider an optical cross-cavity system coupled to a $\Lambda$-type three-level atom. The bright component of an incoming light undergoes reflection caused by a high cooperativity between the atom and the symmetric collective mode of the cavity setup. Meanwhile, the dark component is transmitted since the atom and the antisymmetric collective mode of the cavity setup are uncoupled. The use of a $\Lambda$-type three-level atom allows the device to be turned on or off by controlling the atomic state. However, the device requires only a two-level atom for its operation. Therefore, although we have primarily explored the optical domain and atomic systems~\cite{brekenfeld}, the principles and techniques discussed here can be adapted and extended to solid-state-based systems, such as in circuit \cite{RevModPhys.93.025005} and waveguide \cite{RevModPhys.95.015002} quantum electrodynamics. Our results open new horizons for beam splitters, contributing to the advancement of quantum optics and quantum technologies.

\begin{acknowledgments}
This work was supported by the Brazilian National Council for Scientific and Technological Development (CNPq, Grants Nos.~405712/2023-5, 311612/2021-0 and 465469/2014-0), by CAPES-COFECUB (CAPES, Grants Nos. 88887.711967/2022-00), and by the S\~ao Paulo Research Foundation (FAPESP, Grants Nos. 2019/11999-5, 2018/22402-7, and 2022/00209-6).

\end{acknowledgments}

\appendix

\section{HEISENBERG-LANGEVIN EQUATIONS WITH HOLSTEIN-PRIMAKOFF APPROXIMATION} \label{app:hl}
For the Hamiltonian given in Eq.~\eqref{Hamiltonian}, the Heisenberg-Langevin equation for an arbitrary operator of the system $\mathcal{O}(t)$, which leads to Eqs.~\eqref{adot2}--\eqref{inoutrel}, is written as \cite{qnoise}
\begin{align}\label{iogen}
        \dot{\mathcal{O}} = &-i[\mathcal{O},H_\text{sys}] - \textstyle\sum\nolimits_{l=1}^{2}\Gamma_{l}([\mathcal{O},\sigma^{l}_{+}]\sigma^{l}_{-} - \sigma^{l}_{+}[\mathcal{O},\sigma^{l}_{-}]) \nonumber \\
        &- \textstyle\sum\nolimits_{z=a}^{b} [\mathcal{O},z^{\dagger}](\kappa_{z}^{r}z  -  \sqrt{2\kappa_{z}^{r}}z^{r}_\text{in}) \nonumber \\
        &+ \textstyle\sum\nolimits_{z=a}^{b}(\kappa_{z}^{r} z^\dagger  -  \sqrt{2\kappa_{z}^r}z^{r\dagger}_\text{in})[\mathcal{O},z] \nonumber \\
        &- \textstyle\sum\nolimits_{z=a}^{b} [\mathcal{O},z^{\dagger}](\kappa_{z}^{t}z  +  \sqrt{2\kappa_{z}^{t}}z^{t}_\text{in}) \nonumber \\
        &+ \textstyle\sum\nolimits_{z=a}^{b}(\kappa_{z}^{t} z^\dagger  +  \sqrt{2\kappa_{z}^t}z^{t\dagger}_\text{in})[\mathcal{O},z].
\end{align}
Assuming the Holstein-Primakoff approximation ($\sigma_z  \approx -\mathbf{1}$), the Fourier transform of Eqs.~\eqref{adot2}--\eqref{sdot2} yields, in the Heisenberg picture,
\begin{align}
        \sigma_{-}^{1}(\omega) &= -\frac{i \Tilde{g} \sqrt{2}}{(\Gamma - i\Delta)}X^{+}(\omega), \label{sm} \\
        X^{+}(\omega) &= \frac{\sqrt{2\kappa_r}(\Gamma - i\Delta)}{(\kappa - i\Delta)(\Gamma - i\Delta) + (\Tilde{g}\sqrt{2})^{2}} X_\text{in}^{+}(\omega), \label{xpp} \\
        X^{-}(\omega) &= \frac{\sqrt{2\kappa_r}}{\kappa - i\Delta} X_\text{in}^{-}(\omega), \label{xmm}
\end{align}
with $X^{\pm}(\omega) = [a(\omega)\pm b(\omega)]/\sqrt{2}$, $\kappa = \kappa_r + \kappa_t$, $\Delta = \omega - \omega_c$, and $\Tilde{g} = g\ket{g_1}\bra{g_1}$. The latter is an ad hoc assumption introduced to ensure that $\sigma_{-}^{1}(\omega) = 0$ when the atom is initially in the state $\ket{g_2}$. 

Giving the collective-mode operators for the incoming field, $X_\text{in}^{\pm}(\omega)$, for the reflected field, $X_\text{out}^{\pm}(\omega)$, and for the transmitted one, $Y_\text{out}^{\pm}(\omega)$, we obtain the system response from the input-output relations
\begin{align}\label{eqax21}
        &X_\text{out}^{\pm}(\omega) = x_{\pm}^{*}(\omega) X_\text{in}^{\pm}(\omega),  \\ \label{eqax22}
        &Y_\text{out}^{\pm}(\omega) = y_{\pm}^{*}(\omega) X_\text{in}^{\pm}(\omega),   
    \end{align}
with
    \begin{align}\label{xp}
        &x_{+}^{*}(\omega) \equiv \frac{(\bar{\kappa} + i\Delta)(\Gamma - i\Delta) - (\Tilde{g}\sqrt{2})^{2}}{(\kappa - i\Delta)(\Gamma- i\Delta) + (\Tilde{g}\sqrt{2})^{2}},\\ \label{xm}
        &x_{-}^{*}(\omega) \equiv \left(\frac{\bar{\kappa} + i\Delta}{\kappa - i\Delta}\right),\\ \label{yp}
        &y_{+}^{*}(\omega) \equiv \frac{2\sqrt{\kappa_r \kappa_t}(\Gamma - i\Delta)}{(\kappa - i\Delta)(\Gamma- i\Delta) + (\Tilde{g}\sqrt{2})^{2}},\\ \label{ym}
        &y_{-}^{*}(\omega) \equiv \frac{2\sqrt{\kappa_r \kappa_t}}{\kappa - i\Delta},
    \end{align}
in which $\bar{\kappa} = \kappa_r - \kappa_t$.

\section{SCHRÖDINGER EQUATION FOR THE INITIAL SINGLE-EXCITATION STATE} \label{app:sch}
The single excitation condition in Eq.~\eqref{ISa} yields the following general evolved state 
    \begin{widetext}
     \begin{align}
         \label{ESa}
            \ket{\Psi_t} &= c_{e}(t)\ket{e}\ket{0}_{a}\ket{0}_{b}\ket{0}_{\alpha}^{r}\ket{0}_{\beta}^{r}\ket{0}_{\alpha}^{t}\ket{0}_{\beta}^{t}   \nonumber\\
            &+ \sum^{2}_{k=1}{c^{k}_{a}(t)\ket{g_{k}}\ket{1}_{a}\ket{0}_{b}\ket{0}_{\alpha}^{r}\ket{0}_{\beta}^{r}\ket{0}_{\alpha}^{t}\ket{0}_{\beta}^{t}}    
            + \sum^{2}_{k=1}\sum_{\ell=r,t}{\int^{\infty}_{-\infty}d\omega\xi_{k}^{\ell}(\omega,t)\ket{g_{k}}\ket{0}_{a}\ket{0}_{b}A^{\dagger}_{\ell}(\omega)\ket{0}_{\alpha}^{r}\ket{0}_{\beta}^{r}\ket{0}_{\alpha}^{t}\ket{0}_{\beta}^{t}}  \nonumber\\
            &+ \sum^{2}_{k=1}{c^{k}_{b}(t)\ket{g_{k}}\ket{0}_{a}\ket{1}_{b}\ket{0}_{\alpha}^{r}\ket{0}_{\beta}^{r}\ket{0}_{\alpha}^{t}\ket{0}_{\beta}^{t}}    
            + \sum^{2}_{k=1}\sum_{\ell=r,t}{\int^{\infty}_{-\infty}d\omega\zeta_{k}^{\ell}(\omega,t)\ket{g_{k}}\ket{0}_{a}\ket{0}_{b}B^{\dagger}_{\ell}(\omega)\ket{0}_{\alpha}^{r}\ket{0}_{\beta}^{r}\ket{0}_{\alpha}^{t}\ket{0}_{\beta}^{t}}.
     \end{align}
    \end{widetext} 
Inserting $\ket{\Psi_t}$ into $i\partial_{t}\ket{\Psi_t} = (H - i\Gamma \sigma_{ee})\ket{\Psi_t}$, removing $C(\omega)$ and $D(\omega)$ from $H$, yields the following sets of coupled integro-differential equations for the probability amplitudes
    \begin{align}
        \label{ideq}
        \dot{c}_{e}(t) &= - \Gamma c_{e}(t) - i g_a c_{a}^{1}(t) - i g_b c_{b}^{1}(t),\\
        \dot{c}_{a}^{k}(t) &= -i g_{a}^{*}\delta_{k,1}c_{e}(t) + \sum_{\ell=r,t}\sqrt{\tfrac{\kappa_{a}^{\ell}}{\pi}} \int d\omega \xi_{k}^{\ell}(\omega, t), \\
        \dot{c}_{b}^{k}(t) &= -i g_{b}^{*}\delta_{k,1}c_{e}(t) + \sum_{\ell=r,t}\sqrt{\tfrac{\kappa_{b}^{\ell}}{\pi}} \int d\omega \zeta_{k}^{\ell}(\omega, t),\\
        \dot{\xi}_{k}^{\ell}(\omega, t) &= -\sqrt{\tfrac{\kappa_{a}^{\ell}}{\pi}} c_{a}^{k}(t) -i \omega {\xi}_{k}^{\ell}(\omega, t), \\
        \dot{\zeta}_{k}^{\ell}(\omega, t) &= -\sqrt{\tfrac{\kappa_{b}^{\ell}}{\pi}} c_{b}^{k}(t) -i \omega {\zeta}_{k}^{\ell}(\omega, t).
    \end{align}
By integrating the equations for $\dot{\xi}_{l}$ and $\dot{\zeta}_{l}$ from an initial time $t_{0}$ to $t>t_0$, we obtain
    \begin{align}
        \xi_{k}^{\ell}(\omega,t) &= \xi_{k}^{\ell}(\omega,t_0) e^{-i\omega(t - t_0)}  \nonumber \\ &~~~~- \sqrt{\frac{\kappa_{a}^{\ell}}{\pi}}\int^{t}_{t_0}d\tau c^{k}_{a}(\tau) e^{-i\omega(t - \tau)}, \label{xi0} \\
        \zeta_{k}^{\ell}(\omega,t) &= \zeta_{k}^{\ell}(\omega,t_0) e^{-i\omega(t - t_0)} \nonumber \\ &~~~~- \sqrt{\frac{\kappa_{b}^{\ell}}{\pi}}\int^{t}_{t_0}d\tau c^{k}_{b}(\tau) e^{-i\omega(t - \tau)}, \label{zeta0}
    \end{align}
for $k=1,2$. Now we apply the time limit for these solutions. For a past time $t_0\to-\infty$ we obtain $\xi_{k}^{\ell}(\omega,t_{0})\equiv 
(-1)^{\delta_{t,\ell}} \xi_\text{in}^{k\ell}(\omega)$ and $\zeta_{k}^{\ell}(\omega,t_{0}) \equiv  (-1)^{\delta_{t,\ell}} \zeta_\text{in}^{k\ell}(\omega)$ where the incoming pulse is still at a sufficiently large distance from the cavity, with $\xi_\text{in}^{k t}(\omega) = \zeta_\text{in}^{k t}(\omega) = 0$ since there are no fields coming from these reservoirs in our case. The minus sign comes from the convention that takes into account the propagation direction of the fields in their amplitudes~\cite{walls2008}. Also, we integrate from $t$ to a future time $t_1 > t$, such that
    \begin{align}
        \xi_{k}^{\ell}(\omega,t) &= \xi_{k}^{\ell}(\omega,t_1) e^{-i\omega(t - t_1)} \nonumber \\ &~~~~+ \sqrt{\frac{\kappa_{a}^\ell}{\pi}}\int^{t_{1}}_{t}d\tau c^{k}_{a}(\tau) e^{-i\omega(t - \tau)}, \label{xi1} \\
        \zeta_{k}^{\ell}(\omega,t) &= \zeta_{k}^{\ell}(\omega,t_1) e^{-i\omega(t - t_1)}\nonumber \\ &~~~~ + \sqrt{\frac{\kappa_{b}^\ell}{\pi}}\int^{t_{1}}_{t}d\tau c^{k}_{b}(\tau) e^{-i\omega(t - \tau)}. \label{zeta1}
    \end{align}
For a future time $t_1\to+\infty$, we obtain $\xi_{k}^{\ell}(\omega,t_{1})\equiv (-1)^{\delta_{r,\ell}}\xi_\text{out}^{k\ell}$ and $\zeta_{k}^{\ell}(\omega,t_{1}) \equiv (-1)^{\delta_{r,\ell}}\zeta_\text{out}^{k\ell}$.

The square-normalized temporal shapes of the incoming and outgoing pulses related to cavity $a$ and $b$ are
    \begin{align}
    \label{ftransf}
        \alpha_\text{in}^{k\ell}(t) &= \frac{1}{\sqrt{2\pi}}\int^{+\infty}_{-\infty}\xi_\text{in}^{k\ell}(\omega)e^{-i\omega(t-t_0)}, \\ \beta_\text{in}^{k\ell}(t) &= \frac{1}{\sqrt{2\pi}}\int^{+\infty}_{-\infty}\zeta_\text{in}^{k\ell}(\omega)e^{-i\omega(t-t_0)}, \\ \label{ftransf2}
        \alpha_\text{out}^{k\ell}(t) &= \frac{1}{\sqrt{2\pi}}\int^{+\infty}_{-\infty}\xi_\text{out}^{k\ell}(\omega)e^{-i\omega(t-t_1)}, \\ \beta_\text{out}^{k\ell}(t) &= \frac{1}{\sqrt{2\pi}}\int^{+\infty}_{-\infty}\zeta_\text{out}^{k\ell}(\omega)e^{-i\omega(t-t_1)}.
    \end{align}
As a result, combining the equations above with the relations obtained for $\xi_{k}^{\ell}(\omega,t)$ and $\zeta_{k}^{\ell}(\omega,t)$ from Eqs.~\eqref{xi0}--\eqref{zeta1}, we can obtain the boundary conditions that relate the temporal shapes of field amplitudes outside the cavities to the intracavity fields when the atom is in $\ket{g_k}$: 
    %
    %
%
    \begin{align}
        \label{I-O bc}
           &\text{Cavity $a$} \nonumber \\ 
           &~~~~~~~~~\text{reflected field: } \alpha^{k r}_\text{out}(t)=\sqrt{2\kappa_{a}^r}c_{a}^{k}(t) - \alpha^{k r}_\text{in}(t),  \\ 
            \label{I-O bc2} 
            &~~~~~\text{transmitted field: } \alpha^{k t}_\text{out}(t)=\sqrt{2\kappa_{a}^t}c_{a}^{k}(t), \\ \nonumber \\
            &\text{Cavity $b$: } \nonumber \\
            \label{I-O bc3} 
            &~~~~~~~~~\text{reflected field:  } \beta^{k r}_\text{out}(t)=\sqrt{2\kappa_{b}^r}c_{b}^{k}(t) - \beta^{k r}_\text{in}(t), \\
            \label{I-O bc4} 
            &~~~~~\text{transmitted field:  }\beta^{k t}_\text{out}(t)=\sqrt{2\kappa_{b}^t}c_{b}^{k}(t).
    \end{align}

The last step to build a solvable set of equations is to eliminate the integral terms in $\dot{c}_{a}^{k}(t)$ and $\dot{c}_{b}^{k}(t)$. Let us take as an example the equation for $\dot{c}_{a}^{k}(t)$,
    \begin{widetext}
    \begin{align}
        \dot{c}_{a}^{k}(t) &= -i g_{a}^{*}\delta_{k,1}c_{e}(t) + \sum_{\ell=r,t}\sqrt{\tfrac{\kappa_{a}^{\ell}}{\pi}} \int^{+\infty}_{-\infty} d\omega \,\xi_{k}^{\ell}(\omega, t) \nonumber  \\
        &= -i g_{a}^{*}\delta_{k,1} c_{e} + \sum_{\ell=r,t}\sqrt{\tfrac{\kappa_{a}^{\ell}}{\pi}}\int^{+\infty}_{-\infty}d\omega\left[\xi_{k}^{\ell}(\omega,t_0) e^{-i\omega(t - t_0)} - \sqrt{\frac{\kappa_{a}^{\ell}}{\pi}}\int^{t}_{t_0}d\tau c^{k}_{a}(\tau) e^{-i\omega(t - \tau)}\right]\nonumber   \\
        &= -i g_{a}^{*}\delta_{k,1} c_{e} + \sqrt{\frac{\kappa_{a}^r}{\pi}}\underbrace{\int^{+\infty}_{-\infty} d\omega \xi^{k r}_\text{in}(\omega)e^{-i\omega(t - t_0)}}_{\sqrt{2\pi} \alpha_\text{in}^{k r}(t)} - \frac{(\kappa_{a}^r + \kappa_{a}^t)}{\pi}\underbrace{\int^{t}_{t_0}d\tau c_{a}^{k}(\tau)\underbrace{\int^{+\infty}_{-\infty}d\omega e^{-\omega(t - \tau)}}_{2\pi \delta(t-\tau)}}_{\pi c_{a}^{k}(t)},
    \end{align}
    \end{widetext}
such that
    \begin{equation}
        \dot{c}_{a}^{k}(t) = -ig_{a}^{*}\delta_{k,1}c_{e}  + \sqrt{2\kappa_{a}^r}\alpha_\text{in}^{kr}(t) - (\kappa_{a}^r + \kappa_{a}^t)c^{k}_{a}(t).
    \end{equation}
The same procedure can be taken for $c_{b}^{k}(t)$, yielding the following set of coupled differential equations
    \begin{align}
        \dot{c}_{a}^{1}(t) &=-ig_{a}^{*}c_{e}(t) - (\kappa_{a}^r + \kappa_{a}^t)c^{1}_{a}(t) + \sqrt{2\kappa_{a}^r}\alpha_\text{in}^{1 r}(t),  \\ 
        \dot{c}_{b}^{1}(t) &=-ig_{b}^{*}c_{e}(t) - (\kappa_{b}^r + \kappa_{b}^t)c^{1}_{b}(t) + \sqrt{2\kappa_{b}^r}\beta_\text{in}^{1 r}(t), \\ 
        \dot{c}_{e}(t) &= -\Gamma c_{e}(t) - ig_{a}c_{a}^{1}(t) - i g_{b}c_{b}^{1}(t), \\
        \dot{c}_{a}^{2}(t) &= -(\kappa_{a}^r + \kappa_{a}^t)c^{2}_{a}(t) + \sqrt{2\kappa_{a}^r}\alpha_\text{in}^{2r}(t), \\
        \dot{c}_{b}^{2}(t) &= -(\kappa_{b}^r + \kappa_{b}^t)c^{2}_{b}(t) + \sqrt{2\kappa_{b}^r}\beta_\text{in}^{2r}(t).
    \end{align} 
Given a certain input pulse [$\alpha_\text{in} (t)$ and/or $\beta_\text{in} (t)$] and consider the initial state of Eq. (\ref{ISa}), we have $\alpha^{kr}_\text{in}(t) = \lambda_k \mu_a \alpha_\text{in}(t)$, $\beta^{kr}_\text{in}(t) = \lambda_k \mu_b \beta_\text{in}(t)$ and $c_{a}^{k}(t_0) = c_{b}^{k}(t_0) = c_e(t_0) = 0$. Then, it is straightforward to numerically solve the system dynamics, and hence access the outgoing pulse dynamics through $\alpha^{k\ell}_\text{out}(t)$ and $\beta^{k\ell}_\text{out}(t)$ determined by Eqs.~\eqref{I-O bc}--\eqref{I-O bc4}.

\section{PROBABILITIES OF REACHING THE TARGET STATE} \label{app:prob}

\subsection{HEISENBERG-LANGEVIN PICTURE}
Consider an input state $\ket{g_1}\ket{\psi_\text{in}}$ comprising a generic superposition of single-photon bright and dark states, with
\begin{align}
        \label{psi0}        \ket{\psi_\text{in}} & =\left[\mu_{+}^{\ast} X_\text{in}^{+}(\omega_c) + \mu_{-}^{\ast} X_\text{in}^{-}(\omega_c)\right]^\dagger \ket{0}_{\alpha}^{r} \ket{0}_{\beta}^{r}   \ket{0}_{\alpha}^{t}  \ket{0}_{\beta}^{t} \nonumber \\
         &= \left(\mu_{a}\ket{1_\text{in}}_{\alpha}^{r} \ket{0}_{\beta}^{r}+ \mu_{b}\ket{0}_{\alpha}^{r} \ket{1_\text{in}}_{\beta}^{r}\right)   \ket{0}_{\alpha}^{t}  \ket{0}_{\beta}^{t}.
    \end{align}
In this case, for $C\gg1$ and considering identical and symmetrical cavities, the following target state is expected for the output field
\begin{align}
        \label{psitarget}        \ket{\psi_\text{tgt}} &=\left[-\mu_{+}^{\ast} X_\text{out}^{+}(\omega_c) + \mu_{-}^{\ast} Y_\text{out}^{-}(\omega_c)\right]^\dagger \ket{0}_{\alpha}^{r} \ket{0}_{\beta}^{r}   \ket{0}_{\alpha}^{t}  \ket{0}_{\beta}^{t}.
    \end{align}
The probability of having this happening for any value of $C$ is
\begin{align}
    P_\text{tgt}^\text{HP} &=  \vert\bra{g_1}\langle \psi_\text{tgt} \vert \psi_\text{in} \rangle\ket{g_1} \vert^2 \nonumber \\
    & = \vert \bra{g_1}\bra{0}_{\alpha}^{r} \bra{0}_{\beta}^{r}   \bra{0}_{\alpha}^{t}  \bra{0}_{\beta}^{t} \left[-\mu_{+}^{\ast} X_\text{out}^{+}(\omega_c) + \mu_{-}^{\ast} Y_\text{out}^{-}(\omega_c) \right] \nonumber \\ &~~~~~\times \left[\mu_{+}^{\ast} X_\text{in}^{+}(\omega_c) + \mu_{-}^{\ast} X_\text{in}^{-}(\omega_c)\right]^\dagger\ket{0}_{\alpha}^{r} \ket{0}_{\beta}^{r}   \ket{0}_{\alpha}^{t}  \ket{0}_{\beta}^{t} \ket{g_1} \vert^2 \nonumber \\
    & = \vert - \vert \mu_{+}\vert^2\bra{g_1} x_{+}(\omega_c)\ket{g_1} + \vert \mu_{-}\vert^2\bra{g_1} y_{-}(\omega_c)\ket{g_1} \vert^2 \nonumber \\ 
    &= \left( \vert\mu_{-}\vert^{2} + \vert\mu_{+}\vert^{2}\frac{C}{1+C} \right)^2,
\end{align}
for which we used $[X_\text{out}^{+}(\omega_c),  X_\text{in}^{+}(\omega_c) ^\dagger] = x_{+}(\omega_c)$, $[Y_\text{out}^{-}(\omega_c),  X_\text{in}^{-}(\omega_c) ^\dagger] = x_{-}(\omega_c)$, $[X_\text{out}^{+}(\omega_c),  X_\text{in}^{-}(\omega_c) ^\dagger] =0$, and $ [Y_\text{out}^{-}(\omega_c),  X_\text{in}^{+}(\omega_c) ^\dagger] = 0$.

\subsection{SCHRÖDINGER PICTURE}

Consider
\begin{widetext}
\begin{align} \label{Initsta}
    \ket{\Psi_{t_0}(\omega)}&=\ket{g_1} \ket{0}_{a}  \ket{0}_{b} \left(\mu_{a}\ket{1}_{\alpha}^{r}\ket{0}_{\beta}^{r}+ \mu_{b}\ket{0}_{\alpha}^{r}\ket{1}_{\beta}^{r}\right)  \ket{0}_{\alpha}^{t} \ket{0}_{\beta}^{t} \nonumber \\
    &=\ket{g_1} \ket{0}_{a}  \ket{0}_{b}  \int d\omega \xi_\text{in}(\omega)\underbrace{\left(\mu_{+}\ket{\Psi_\mathcal{B}^1(\omega)}_r+ \mu_{-}\ket{\Psi_\mathcal{D}^1(\omega)}_r\right)  \ket{0}_{\alpha}^{t} \ket{0}_{\beta}^{t}}_{\ket{\psi_\text{in} (\omega)}}.
\end{align}
\end{widetext}
For identical and symmetrical cavities, the reservoirs are expected to be found in the following final state in a high-cooperativity regime
\begin{align} \label{finalstate2}
   \ket{\psi_\text{tgt}(\omega)}=-\mu_{+}\ket{\Psi_\mathcal{B}^1(\omega)}_r\ket{0}_{\alpha}^{t} \ket{0}_{\beta}^{t}+ \mu_{-}\ket{0}_{\alpha}^{r} \ket{0}_{\beta}^{r}\ket{\Psi_\mathcal{D}^1(\omega)}_t.
\end{align}
The probability of this happening is given by

\begin{align}
    P_\text{tgt} &= \int^{+\infty}_{-\infty} \!\!\!\! d\omega \,\abs{\bra{g_1} \bra{0}_a \bra{0}_b \langle{\psi_\text{tgt}(\omega)} \ket{\Psi_t} \rvert^2 \nonumber \\
    &= \int^{+\infty}_{-\infty} \!\!\!\! dt \, \lvert \mu_{-} [\alpha^{1 t}_\text{out}(t) - \beta^{1 t}_\text{out}(t)] - \mu_{+}[\alpha^{1 r}_\text{out}(t) + \beta^{1 r}_\text{out}(t)] }^2.
\end{align}

\bibliography{refs}

\begin{thebibliography}{42}
\bibitem [{sig()}]{sign}%
  \BibitemOpen
  \href@noop {} {\bibinfo {title} {Here there is a phase convention between the incoming and outgoing external fields, where their propagation directions are explicitly incorporated in their amplitudes.}}\BibitemShut {Stop}%
\end{thebibliography}

\end{document}